\documentclass[]{aa}
\usepackage{txfonts}
\usepackage{graphicx}
\begin{document}

\title{The nearby eclipsing stellar system $\delta$ Velorum}
\subtitle{I. Origin of the infrared excess from VISIR and NACO imaging
\thanks{Based on observations made with ESO telescopes at the La Silla Paranal Observatory,
under ESO programs 081.D-0109(B) and 081.D-0109(C).}}
\titlerunning{The eclipsing stellar system $\delta$ Velorum - I. VISIR and NACO imaging}
\authorrunning{P. Kervella et al.}
\author{
P. Kervella\inst{1}
\and
F. Th\'evenin\inst{2}
\and
M. G. Petr-Gotzens\inst{3}
}
\offprints{P. Kervella}
\mail{Pierre.Kervella@obspm.fr}
\institute{
LESIA, Observatoire de Paris, CNRS\,UMR\,8109, UPMC, Universit\'e Paris Diderot, 5 Place Jules Janssen, 92195 Meudon, France
\and
Universit\'e de Nice-Sophia Antipolis, Lab. Cassiop\'ee, UMR\,6202, Observatoire de la C\^ote d'Azur, BP 4229, 06304 Nice, France
\and
European Southern Observatory, Karl-Schwarzschild-Str. 2,
D-85748 Garching, Germany
}
\date{Received ; Accepted}
\abstract
% Context
{The triple stellar system $\delta$\,Vel system presents a significant infrared excess, whose origin is still being debated.
A large infrared bow shock has been discovered using \emph{Spitzer}/MIPS observations.
Although it appears as a significant contributor to
the measured IR excess, the possibility exists that a circumstellar IR excess is present
around the stars of the system.}
% Aims
{The objective of the present VISIR and NACO observations is to identify whether one of the stars of the
$\delta$\,Vel system presents a circumstellar photometric excess in the thermal IR domain and to quantify it.}
% Methods
{We observed $\delta$\,Vel using the imaging modes of the ESO/VLT instruments VISIR (in BURST mode)
and NACO to resolve the A--B system (0.6\arcsec\ separation) and obtain the photometry of each star.
We also obtained one NACO photometry epoch precisely at the primary (annular) eclipse of $\delta$\,Vel~Aa by Ab.}
% Results
{Our photometric measurements with NACO (2.17\,$\mu$m), complemented by the existing
visible photometry allowed us to reconstruct the spectral energy distribution of the three stars.
We then compared the VISIR photometry (8.6--12.8\,$\mu$m) to the expected photospheric emission
from the three stars at the corresponding wavelengths.}
% Conclusions
{We can exclude the presence of a circumstellar thermal infrared excess around $\delta$\,Vel A or B down to a few percent level. This supports the conclusions of G\'asp\'ar et al.~(2008) that the IR excess of $\delta$\,Vel has an interstellar origin, although a cold circumstellar disk could still be present. In addition, we derive the spectral types of the three stars Aa, Ab, and B (respectively A2IV, A4V and F8V), and we estimate the age of the system around 400--500\,Myr.}
\keywords{Stars: individual: (HD 74956, $\delta$ Vel); Stars: binaries: eclipsing; Methods: observational; Techniques: high angular resolution}

\maketitle

%__________________________________Introduction
\section{Introduction}

The southern star \object{$\delta$ Vel} (\object{HD 74956}, \object{HIP 41913}, \object{GJ 321.3}, \object{GJ 9278}) is the 43$^{\rm rd}$ brightest star in the visible sky, with $m_V = 1.96$, and is located at a distance of 24.4\,pc ($\pi = 40.90 \pm 0.38$\,mas; ESA~\cite{esa97}). Amazingly, $\delta$\,Vel was discovered only very recently by Otero et al.~(\cite{otero00}) to host the brightest eclipsing binary system in the southern sky, one of the very few observable with the unaided eye.
Its orbital period of 45\,days is exceptionally long in terms of probability of occurrence, especially for such a nearby star.
A historical note on the discovery and recent developments in the study of $\delta$\,Vel can be found in Argyle, Alzner \& Horch~(\cite{argyle02}), together with the orbital parameters of the A--B system (orbital period of $\approx 142$\,years). In the following, we refer to the eclipsing pair as $\delta$\,Vel~A, and to the fainter, visual component as $\delta$\,Vel~B. The two eclipsing components of A are referred to as Aa and Ab, by order of decreasing brightness. Kellerer et al.~(\cite{kellerer07}) excluded the physical association of the angularly nearby pair of stars sometimes labeled $\delta$\,Vel~C and D with the $\delta$\,Vel~A--B system.

An infrared (IR) excess associated with the $\delta$\,Vel~A--B system had been
detected by IRAS~(\cite{ipac86}). This indicated that $\delta$\,Vel could belong to 
the ''Vega-like'' class of objects,
i.e.\ a main sequence star surrounded by a optically thin debris disk. However,  
G\'asp\'ar et al.~(\cite{gaspar08}) recently discovered a spectacular infrared (IR) bow shock around $\delta$\,Vel at 24 and 70\,$\mu$m using \emph{Spitzer}/MIPS images. This very large structure ($\approx1\arcmin$) is explained by these authors as the result of the heating and compression of the interstellar medium by the photons from $\delta$\,Vel, as the trio moves through the interstellar medium (ISM). From a detailed modeling of the star--ISM interaction, G\'asp\'ar et al. conclude that the bow shock contribution is sufficient to explain the observed IR excess of $\delta$\,Vel without resorting to a circumstellar debris disk. They also note that accretion of interstellar material ($\lambda$~Bootis phenomenon) could nevertheless take place on $\delta$\,Vel, although at a slow rate.

In this article, we present the results of our search for circumstellar IR excess in the inner $\delta$\,Vel~A-B system. Sect.~\ref{observations} is dedicated to the description of our high angular resolution VISIR and NACO observations, whose analysis is presented in Sect.~\ref{analysis}. In Sect.~\ref{photometric_analysis}, we compute the magnitudes of each of the three stars of the system, that are used to derive their physical parameters. We present in Sect.~\ref{discussion} our estimates of the thermal IR excesses of $\delta$\,Vel Aab and B and a short analysis of the evolutionary state of the system. 

%__________________________________Observations
\section{Observations and data reduction\label{observations}}

\subsection{VISIR}

For the thermal-IR part of our program, we used the VISIR instrument (Lagage et al.~\cite{lagage04}), installed at the Cassegrain focus of the Melipal telescope (UT3) of the ESO/Very Large Telescope (Paranal, Chile). VISIR is a mid-IR imager, that also provides a slit spectrometer. VISIR can in principle reach a very high spatial resolving power, thanks to the 8\,m diameter of the telescope. However, under standard conditions at Paranal (median seeing of 0.8$\arcsec$ at 0.5\,$\mu$m), the 8\,m telescope is not diffraction limited in the MIR (seeing $\approx 0.4\arcsec$ vs. 0.3$\arcsec$ diffraction). A few moving speckles and tip-tilt usually degrade the quality of the image (see e.g. Tokovinin, Sarazin \& Smette~\cite{tokovinin07}). To overcome this limitation, a specific mode of the instrument, the BURST mode (Doucet et al.~\cite{doucet07a}, \cite{doucet07b}), give the possibility to record separately a large number (several tens of thousand) very short exposures ($\Delta t \lesssim 50$\,ms), in order to freeze the turbulence. The data processing procedure we developed to reduce the resulting data cubes is described in Kervella \& Domiciano de Souza~(\cite{kervella07}). During the processing, the frames were precisely co-aligned on the position of $\delta$\,Vel~A.

\begin{figure}[h!]
\centering
\includegraphics[width=2.9cm]{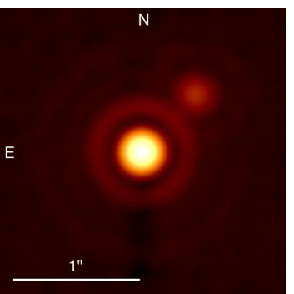}
\includegraphics[width=2.9cm]{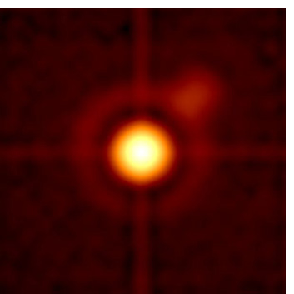}
\includegraphics[width=2.9cm]{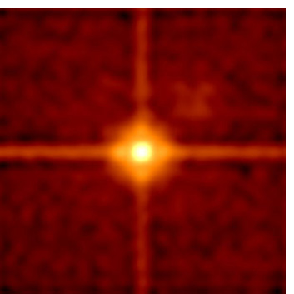}
\includegraphics[width=2.9cm]{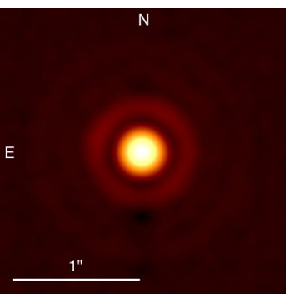}
\includegraphics[width=2.9cm]{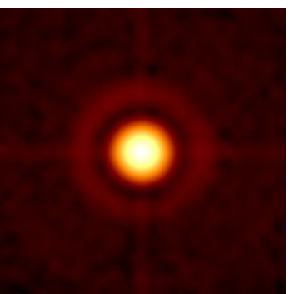}
\includegraphics[width=2.9cm]{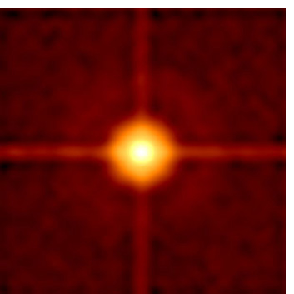}
\caption{{\it Top row:} VISIR average images of $\delta$\,Vel in the PAH1 (\#A, B, E and F in Table~\ref{visir_log}), PAH2 (\#G, H, I) and NeII (\#L) bands. {\it Bottom row:} Calibrator images in the PAH1 (HD\,67582), PAH2 (HD\,67582) and NeII (HD\,80007) bands. \label{avg-images}}
\end{figure}

We observed $\delta$\,Vel and its three calibrators, \object{HD 82668}, \object{HD 67582} and \object{HD 80007} in visitor mode at Paranal during the night of 23-24 April 2008. At that time $\delta$\,Vel was out of ecplipse. The sequence of the observations is presented in Table~\ref{visir_log}. We adopted standard chopping and nodding offsets of $8\arcsec$, with respective periods of 4 and 90\,s. The calibrators were selected from the Cohen et al.~(\cite{cohen99}) catalog of spectrophotometric standards for IR wavelengths, except \object{HD\,80007} (for the NeII filter).  For each observation of $\delta$\,Vel and its calibrators, we selected during the data processing $\approx 40$\% of the total number of frames, rejecting those with the lowest Strehl ratio (estimated from the peak intensity in the frame). For instance, the first two observations $\#A$ and $B$ of $\delta$\,Vel (20\,000 selected frames) correspond to a total of 50\,000 frames before selection. The number of frames listed in Table~\ref{visir_log} corresponds to the result of this selection. 
The cubes were then averaged to obtain diffraction-limited images in our three filters: PAH1, PAH2, and NeII. that have respective central wavelengths of $\lambda = 8.59$, 11.25 and 12.81\,$\mu$m. The resulting images of $\delta$\,Vel and the calibrators are presented in Fig.~\ref{avg-images}. The observation $\#C$ of HD\,82668 was affected by saturation of the array, and was not used for our anaysis.

\begin{table}
\caption{Log of the observations of $\delta$\,Vel and its calibrators,  \object{HD 82668}, \object{HD 67582} and \object{HD 80007} with VISIR. MJD is the modified Julian date of the middle of the exposures on the target, minus 54\,580. The Detector Integration Time (DIT) is given in milliseconds for one BURST image. $\theta$ is the seeing in the visible ($\lambda=0.5\,\mu$m) as measured by the observatory DIMM sensor, in arcseconds. $N$ exp. is the number of selected and averaged exposures. The airmass (AM) is the average airmass of the observation.} 
\label{visir_log}
\begin{tabular}{cccccccc}
\hline \hline
\# & MJD$^*$ & Star & Filter & DIT & $N$ exp. & $\theta$\,($\arcsec$) & AM \\
\hline
A & 0.0134 & $\delta$\,Vel & PAH1 & 20 & 10\,000 & 0.86 & 1.18 \\
B & 0.0231 & $\delta$\,Vel & PAH1 & 20 & 10\,000 & 1.00 & 1.20 \\
C & 0.0374 & HD\,82668 & PAH1 & 20 & 5\,000 & 1.10 & 1.20 \\
D & 0.0603 & HD\,67582 & PAH1 & 16 & 10\,000 & 0.84 & 1.30 \\
E & 0.0774 & $\delta$\,Vel & PAH1 & 16 & 8\,000 & 0.70 & 1.33 \\
F & 0.0874 & $\delta$\,Vel & PAH1 & 16 & 8\,000 & 0.67 & 1.37 \\
G & 0.0977 & $\delta$\,Vel & PAH2 & 8 & 10\,000 & 0.60 & 1.42 \\
H & 0.1093 & $\delta$\,Vel & PAH2 & 8 & 12\,000 & 0.53 & 1.48 \\
I & 0.1206 & $\delta$\,Vel & PAH2 & 8 & 10\,000 & 0.80 & 1.54 \\
J & 0.1363 & HD\,67582 & PAH2 & 8 & 10\,000 & 1.15 & 1.92 \\
K & 0.1480 & HD\,67582 & PAH2 & 8 & 10\,000 & 1.50 & 2.09 \\
L & 0.1632 & $\delta$\,Vel & NeII & 16 & 10\,000 & 1.29 & 1.91 \\
M & 0.1860 & HD\,80007 & NeII & 16 & 10\,000 & 1.12 & 1.46 \\
N & 0.1979 & HD\,80007 & NeII & 16 & 10\,000 & 1.00 & 1.52 \\
\hline
\end{tabular}
\end{table}

%______________ Figure

\subsection{NACO}

We observed $\delta$\,Vel at ten epochs in April-May 2008 using the Nasmyth Adaptive Optics System (NAOS, Rousset et al.~\cite{rousset03}) of the Very Large Telescope (VLT), coupled to the CONICA infrared camera (Lenzen et al.~\cite{lenzen98}), abbreviated as NACO. These observations were obtained to provide high-precision differential astrometry of the eclipsing system $\delta$\,Vel A relative to B, and we will analyze them along this line in a forthcoming article.

We selected the smallest available pixel scale of $13.26 \pm 0.03$\,mas/pix (Masciadri et al.~\cite{masciadri03}), giving a field of view of 13.6$\arcsec$$\times$13.6$\arcsec$. Due to the brightness of $\delta$\,Vel, we employed a narrow-band filter at a wavelength of $2.166 \pm 0.023\,\mu$m (hereafter abbreviated as 2.17) together with a neutral density filter (labeled ``{\tt ND2\_short}"), with a transmission of about 1.5\%.

\begin{figure}[]
\centering
\includegraphics[width=8.7cm]{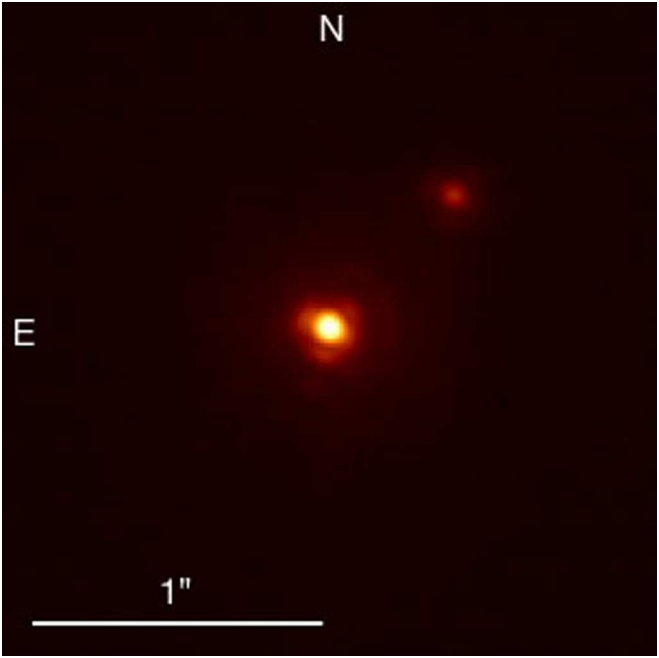}
\caption{Median NACO image of the 51 frames of $\delta$\,Vel obtained on 2008-04-01.\label{naco-image}}
\end{figure}

Table~\ref{naco_log} gives the list of the observations. Each of our ten epochs consisted in approximately 50 short exposures grouped over less than 10 minutes, with a detector integration time (DIT) of 0.8\,s each to avoid saturation. Four such short exposures were co-averaged providing 3.2\,s integrated exposure time per image. A few images (9 over 510 in total) were affected by the opening of the adaptive optics loop, and were removed from the processing. We obtained nine epochs outside of the eclipses, and one precisely at the phase of the primary eclipse (on 18 May 2008), only 23 minutes before the center of the eclipse. As a remark, the {\it primary} eclipse (the deeper in photometry) is when the smaller, cooler star (Ab) passes in front of the bigger, hotter star (Aa), causing an annular eclipse. The {\it secondary} eclipse (shallower in photometry) is when the cooler star is totally eclipsed by the hotter one. The epochs of the primary and secondary eclipses of $\delta$\,Vel~A are given by the ephemeris of Otero\footnote{http://ar.geocities.com/varsao/delta\_Velorum.htm}:
\begin{equation}\label{ephem1}
{\rm Min\,I} = {\rm HJD}\,2452798.557 + 45.1501\ E
\end{equation}
\begin{equation}\label{ephem2}
{\rm Min\,II} = {\rm HJD}\,2452818.200 + 45.1501\ E
\end{equation}
With $E$ the number of elapsed orbits. To retrieve the phases listed in Table~\ref{naco_log}, we converted the modified Julian dates of our observations into heliocentric Julian dates using the tools by Dan Bruton\footnote{http://www.physics.sfasu.edu/astro/javascript/hjd.html}.
The raw images were processed using the Yorick\footnote{http://yorick.sourceforge.net/} and IRAF\footnote{IRAF is distributed by the NOAO, which are operated by the Association of Universities for Research in Astronomy, Inc., under cooperative agreement with the National Science Foundation.} software packages in a standard way. A sample image of $\delta$\,Vel is presented in Fig.~\ref{naco-image}.

\begin{table}
\caption{Log of the observations of $\delta$\,Vel with NACO in the NB2.17 filter. MJD$^*$ is the average modified Julian date minus 54\,000, HJD$^*$ the heliocentric Julian date minus 2\,454\,000 and $\phi$ is the phase of the eclipsing binary (see text). $N$ is the number of frames. As in Table~\ref{visir_log}, $\theta$ is the seeing in the visible and AM is the airmass. The listed figures are median values over the observations.}
\label{naco_log}
\begin{tabular}{ccccccc}
\hline \hline
\# & MJD$^{*}$  & HJD$^{*}$ & $\phi$(Min\,I)  & $N$ exp. & $\theta$\,($\arcsec$) & AM \\
\hline
1 & 557.0224 & 557.5245 & 0.9582 & 51 & 0.80 & 1.157 \\
2 & 560.9976 & 561.4996 & 0.0463 & 46 & 0.91 & 1.164 \\
3 & 562.0121 & 562.5141 & 0.0687 & 51 & 0.67 & 1.156 \\
4 & 563.0048 & 563.5067 & 0.0907 & 51 & 0.89 & 1.158 \\
5 & 576.9715 & 577.4732 & 0.4000 & 51 & 0.66 & 1.156 \\
6 & 579.0231 & 579.5247 & 0.4455 & 50 & 1.06 & 1.191 \\
7 & 580.9917 & 581.4932 & 0.4891 & 51 & 1.24 & 1.164 \\
8 & 591.9748 & 592.4761 & 0.7323 & 51 & 0.59 & 1.175 \\
9 & 593.9732 & 594.4744 & 0.7766 & 50 & 0.77 & 1.180 \\
10 & 604.0442 & 604.5450 & 0.9996 & 49 & 0.76 & 1.479 \\
\hline
\end{tabular}
\end{table}

%__________________________________Image analysis
\section{Image analysis \label{analysis}}

\subsection{VISIR}

%______________ Figure
\begin{figure}[]
\centering
\includegraphics[width=4.4cm]{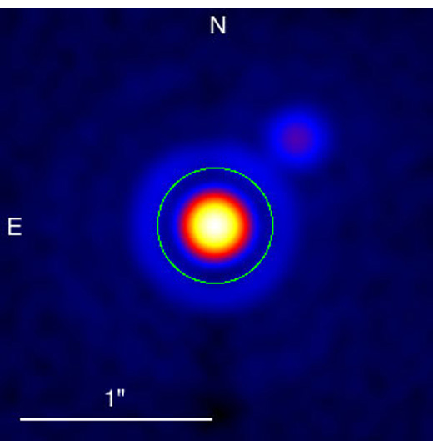}
\includegraphics[width=4.4cm]{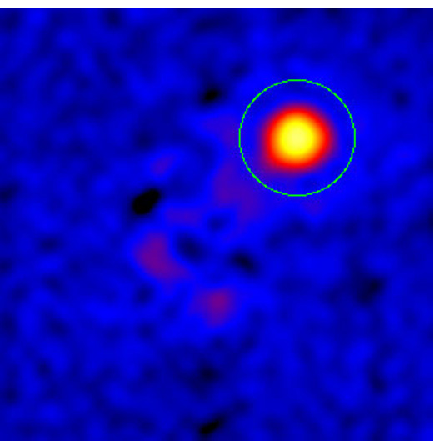}
\caption{Apertures (circles) used for the VISIR photometry of $\delta$\,Vel~A (left) and B (right) shown respectively on the PAH1 image \#B and the subtracted image \#B - \#D (see text for details). \label{apertures}}
\end{figure}

We obtained photometry of $\delta$\,Vel~A in the PAH1 and PAH2 bands using a circular aperture of 0.3$\arcsec$ in radius, centered on the star position (Fig.~\ref{apertures}, left). This small radius is intended to avoid the contamination of the photometry of A by component B.
For $\delta$\,Vel~B, we first subtracted a scaled version of the calibrator star at the position of A from the image before measuring the flux through aperture photometry (Fig.~\ref{apertures}, right). We thus removed the contribution from the diffraction pattern (Airy rings) of $\delta$\,Vel~A at the location of B. This process left almost no residual thanks to the stability of the PSF.

We checked that the VISIR images of $\delta$\,Vel and its calibrators present no detectable diffuse background.
% (sensitivity $\approx$100\,mJy/arcsec$^2$).
This verification was achieved by measuring the average flux at distances larger than 1$\arcsec$ from the central stars, to avoid the inclusion of the high-order Airy rings.  Based on this result, we did not apply a sky subtraction to our observations, and as we used the same apertures for the photometry of $\delta$\,Vel and the flux calibrators, no aperture correction was required either. This non-detection of the diffuse background was expected for two reasons: firstly due to the chopping-nodding observing procedure that removes most of the large scale galactic IR background, and secondly because the sensitivity of our images is insufficient to detect the spatially extended flux from the bow shock in the VISIR bands. The total flux from the model proposed by G\'asp\'ar et al.~(\cite{gaspar08}) is at most 3-4\,mJy in this wavelength range (Sect.~\ref{discussion}), spread over a surface of several arcmin$^2$, far beyond our sensitivity limit.

The integrated fluxes are affected by different atmospheric absorption due to the different airmasses compared to the calibrators. To take this into account, we used the empirical formula by Sch\"utz \& Sterzik~(\cite{schutz05}), that gives the multiplicative correction $C(\lambda,{\rm AM})$ to apply to the raw flux to remove the atmospheric absorption:
\begin{equation}
C(\lambda,{\rm AM}) = 1 + \left[ 0.220 - \frac{0.104}{3}\,\left(\lambda - 8.6\,\mu{\rm m}\right) \right]\,({\rm AM} - 1)
\end{equation}
We corrected separately the different observations of $\delta$\,Vel and its calibrators.

In order to absolutely calibrate the measurements, the PAH1 and PAH2 fluxes from the calibrator HD\,67582 were retrieved from the spectral template of Cohen et al.~(\cite{cohen99}) at the central wavelength of our filters.
For \object{HD\,80007} (\object{HIP 45238}), we adjusted a model spectrum from Castelli \& Kurucz~(\cite{castelli03}) to all the photometry available in the literature to retrieve its irradiance in the NeII filter.
We obtained 5.62\,Jy ($1.02\ 10^{-13}$\,W/m$^2$/$\mu$m) that is within 7\% of the value published on the VISIR instrument web page\footnote{http://www.eso.org/sci/facilities/paranal/instruments/visir/} (5.26\,Jy).
The statistical and calibration uncertainties of the measurements were estimated from the dispersion of the different available exposures (e.g. \#ABEF for the PAH1 filter). The uncertainty on the NeII measurement, for which only one image is available, was taken as the dispersion of the calibrated flux for apertures of 150, 300, 600 and 1200\,mas. The resulting irradiances are presented in Table~\ref{photom_delvel}.

The ratios $\rho = f(B) / f(A)$ of the fluxes of $\delta$\,Vel~B and $\delta$\,Vel~A in each band can be estimated more accurately than the absolute flux of each star due to the removal of the calibration uncertainty. We obtain the following values in the three filters:
\begin{equation}
\rho_{\rm \ 8.6\,\mu m} =  10.3 \pm 0.2\,\%,
\end{equation}
\begin{equation}
\rho_{\rm 11.25\,\mu m} =  11.1 \pm 0.4\,\%,
\end{equation}
\begin{equation}
\rho_{\rm 12.81\,\mu m} =  8.4 \pm 2.3\,\%.
\end{equation}

%___________________Table
\begin{table}
\caption{Measured thermal IR irradiances of $\delta$\,Vel~A and B in S.I. units and in Jy. The relative uncertainty $\sigma$ on the irradiance values is listed in the last column.}
\label{photom_delvel}
\begin{tabular}{lccrcc}
\hline \hline
Star & Filter & $\lambda$ ($\mu$m) & $10^{-14}$ W/m$^2$/$\mu$m & Jy & $\sigma$ \\
\hline
\noalign{\smallskip}
%3,693E-13	2,68E-14	9,10	0,66
%1,238E-13	1,22E-14	5,24	0,52
%6,702E-14	7,44E-15	3,68	0,41
A & PAH1 & 8.59 & $36.93 \pm 2.68$ & $9.10 \pm 0.66$ & 7.3\% \\
A & PAH2 & 11.25 & $12.38 \pm 1.22$ & $5.24 \pm 0.52$ & 9.9\% \\
A & NeII & 12.81& $6.70 \pm 0.74$ & $3.68 \pm 0.41$ & 11.1\% \\
\hline
\noalign{\smallskip}
%3,818E-14	2,81E-15	0,943	0,069	7,4%
%1,378E-14	1,43E-15	0,576	0,060	10,4%
%5,604E-15	1,93E-15	0,307	0,106	34%
B & PAH1 & 8.59 & $3.82 \pm 0.28$ & $0.94 \pm 0.07$ & 7.4\% \\
B & PAH2 & 11.25 & $1.38 \pm 0.14$ & $0.58 \pm 0.06$ & 10.4\%  \\
B & NeII & 12.81& $0.56 \pm 0.19$ & $0.31 \pm 0.11$ & 34\% \\
 \hline
\end{tabular}
\end{table}

\subsection{NACO \label{nacophot}}

\begin{figure}[]
\centering
\includegraphics[width=4.4cm]{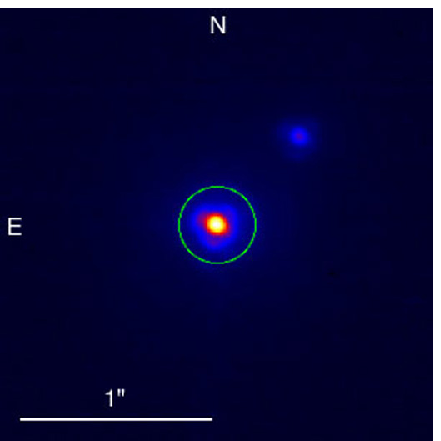}
\includegraphics[width=4.4cm]{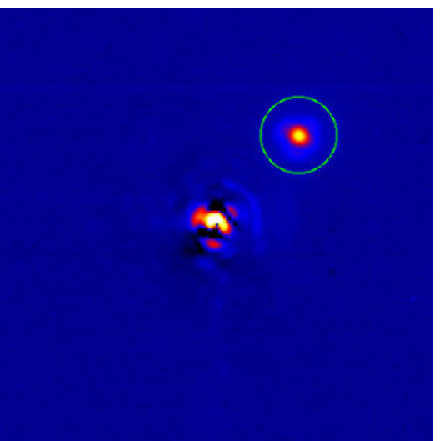}
\caption{Aperture (15\,pixels = 0.2\arcsec in radius) used for the integration of the flux of $\delta$\,Vel~A (left) and B (right), the latter on the ring-median subtracted NACO image (see text for details).\label{naco-subimage}}
\end{figure}

We first obtained classical aperture photometry of $\delta$\,Vel~A using circular apertures of 1 to 40 pixels in radius (single pixel to 0.53\arcsec radius), for each of our ten epochs. Although the two components of $\delta$\,Vel are well separated in the NACO images, the diffuse halo from the residuals of the adaptive optics correction of $\delta$\,Vel~A are not negligible at the position of B. For this reason, the photometry of B cannot be measured directly on the images. To remove the diffuse wings of the point spread function (PSF) of A, we subtracted from each point of the image its ring median at the corresponding radius from $\delta$\,Vel~A. With this procedure, we cleanly subtracted the diffuse background from A, mostly without introducing additional noise. Very little residuals from star A are present on the subtracted image, thanks to the good circularity of the PSF (Fig.~\ref{naco-subimage}). We then measured aperture photometry using the same 1-40\,pixels apertures as for A, centered on B, and computed the flux ratio $\rho = f(B) / f(A)$ of the pair. We do not need to correct for the airmass, as both A and B are affected by the same atmospheric absorption. We checked in the images that the sky background is negligible due to the short exposures and narrow-band filter.

\begin{figure}[]
\centering
\includegraphics[bb=0 0 360 144, width=8.7cm, angle=0]{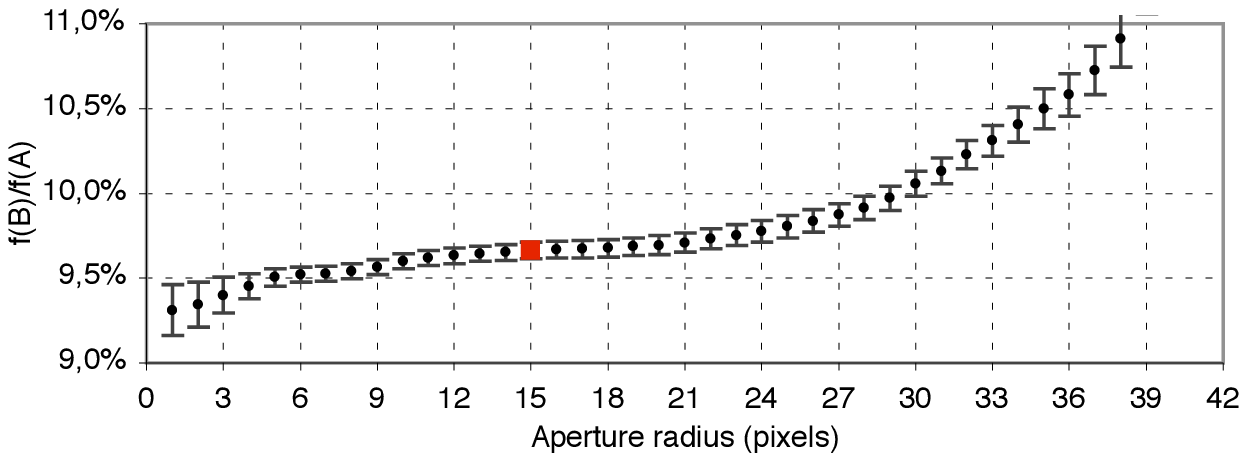}
\caption{Ratio $\rho({\rm 2.17}) = f(B)/f(A)$ at $2.17\,\mu$m as a function of the aperture radius (in pixels) used for the integration of the flux of A and B. An aperture of 15\,pixels was selected for the analysis (square).  \label{naco-aperture}}
\end{figure}

The choice of the aperture radius is an important factor for the accuracy of the flux ratio measurement. In order to obtain the best match between the two stars, it is essential to have the same integration radius for both stars. Over the 0.6\arcsec distance between the two stars, the variation of the PSF shape and Strehl ratio is negligible, especially for the good seeing conditions of our observations. Fig.~\ref{naco-aperture} shows the variation of the derived flux ratio as a function of the aperture radius. There is a clear inflexion point around 15\,pixels in radius (0.2\arcsec). This corresponds to the radius at which the wings of the PSF of B reach the noise level. The increase of $\rho$ observed for larger apertures corresponds to the inclusion of the residuals from A in the photometry.

To estimate the error bar of the flux ratio, we added quadratically the uncertainties due to the choice of aperture radius over a 10-15\,pixel range (considered as systematic, thus not averaging out), and the dispersion of the measurements over the first nine epochs (statistical). The flux ratio is assumed to be constant outside of the eclipses, in order to average the nine measurements (but see also Sect.~\ref{eclipse2}). For the epoch of the eclipse, we considered as uncertainty the standard deviation of the ratios obtained at the other epochs. 

From this procedure, we obtain outside of the eclipse:
\begin{equation}
\rho_{\rm 2.17\,\mu m} = 9.66 \pm 0.05\%
\end{equation}
corresponding to a magnitude difference of $\Delta m = 2.537 \pm 0.005$.
During the primary eclipse of $\delta$\,Vel~A, the flux ratio becomes:
\begin{equation}
\rho_{\rm 2.17\,\mu m}({\rm Eclipse}) = 14.50 \pm 0.12\%
\end{equation}
giving $\Delta m({\rm Eclipse}) = 2.097 \pm 0.009$.
The eclipse depth at $2.17\,\mu$m is therefore $d(\Delta m) = 0.440\pm 0.011$\,mag.
As a remark, we did not deconvolve our NACO images, as we did not obtain a PSF calibrator for this purpose, and we leave the astrometric analysis of these images for a future article.

\begin{figure}[]
\centering
\includegraphics[bb=0 0 360 144, width=8.7cm, angle=0]{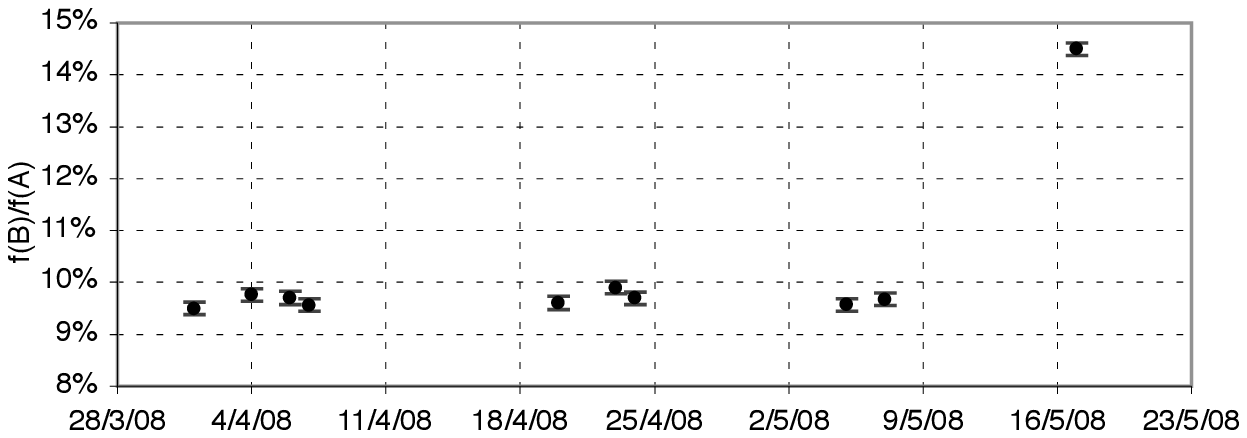}
\caption{Measured value of $f(B)/f(A)$ at $2.17\,\mu$m for our 10 measurement epochs. Note the primary eclipse occuring at the last epoch. \label{naco-photom}}
\end{figure}

\section{Photometric analysis \label{photometric_analysis}}

\subsection{Visible magnitudes}

If not specified otherwise, we assume here that the photometric measurements of the $\delta$\,Vel system available in the literature always include the A and B components, as their angular separation has been smaller than $2\arcsec$ for at least five decades. Bessell~(\cite{bessell90}) gives:
\begin{equation}
 m_V ({\rm Aab+B}) = 1.96 \pm 0.02
 \end{equation}
%and the following colors: $(B-V) = 0.04$, $(V-R_c) = 0.025$, $(R_c-I_c) = 0.030$ and $(V-I_c) = 0.055$,
with an associated uncertainty of $\pm 0.02$\,mag.

The photometric observations from {\it Hipparcos} (ESA~\cite{esa97}) give the differential magnitude of Aab (combined) and B in the $H_p$ band, that are sufficiently close to $V$ to consider the values identical within their error bars:
\begin{equation}
\Delta m_{Hp} ({\rm Aab / B}) \approx \Delta m_{V} ({\rm Aab / B}) = 3.58 \pm 0.07
\end{equation}
Considering Bessel's data, this translates immediately into the following magnitude for Aab and B:
\begin{equation}
 m_V ({\rm Aab}) = 2.00 \pm 0.02
 \end{equation}
\begin{equation}
 m_V ({\rm B}) = 5.54 \pm 0.08
 \end{equation}

In addition, the combined visible photometry available in the literature (see the compilation by Otero et al.~\cite{otero00}) gives us the combined Aab+B magnitudes, the depth of the primary (I) and secondary (II) eclipses:
\begin{equation}
\Delta m_V ({\rm I, Aab+B}) = 0.51 \pm 0.01
\end{equation}
\begin{equation}
\Delta m_V ({\rm II, Aab+B}) = 0.32 \pm 0.01
\end{equation}
By removing the flux contribution of B computed precedently, we obtain the following eclipse depths for Aab only:
\begin{equation}
\Delta m_V ({\rm I, Aab}) = 0.53 \pm 0.01
\end{equation}
\begin{equation}
\Delta m_V ({\rm II, Aab}) = 0.33 \pm 0.01
\end{equation}
As shown by the curve presented by Otero et al.~(\cite{otero00}), the secondary eclipse of $\delta$\,Vel~Aab is total, and we can thus derive immediately the visible magnitudes of Aa and Ab:
\begin{equation}
 m_V ({\rm Aa}) = 2.33 \pm 0.03
 \end{equation}
\begin{equation}
 m_V ({\rm Ab}) = 3.44 \pm 0.03
 \end{equation}

\subsection{Infrared magnitudes}

In addition, our NACO observations (Sect.~\ref{nacophot}) give us the relative fluxes of Aab and B in the narrow band filter band at $2.17\,\mu$m with a FWHM of $0.023\,\mu$m: $\rho_{\rm 2.17\,\mu m} = 9.66 \pm 0.05\%$. As this wavelength corresponds to an absorption line of hydrogen that is present in the spectra of hot stars, the conversion of the measured narrow-band ratios to standard $K$ band ratios requires taking the position of the quasi-monochromatic wavelengths within the bands and the shape of the observed spectra into account. 

We thus proceeded in two steps to obtain the conversion between the narrow-band flux ratio and the broadband flux ratio. Starting from the narrow-band fluxes measured on $\delta$\,Vel's components, we computed the radius and effective temperature of each star assuming (as a first approximation) that the $K$ flux ratios are identical to the $2.17\,\mu$m ratios, using the method descussed in Sect.~\ref{delvel_params}. This gave us a first estimation of the spectral types of the stars. We then used the Pickles~(\cite{pickles98})\footnote{http://www.ifa.hawaii.edu/users/pickles/AJP/hilib.html} reference spectra corresponding to their spectral types (A2IV, A4V and F8V) to recover the corresponding broadband flux ratio. The $K$ band standard filter profile was taken from Bessell \& Brett~(\cite{bessell88}). We observed that the true ratios in the $K$ band are very close to the ratios in the $2.17\,\mu$m filter, within 0.01\,mag for all three stars, which is small compared to the other uncertainties. We thus neglect this difference in the following analysis.

The COBE/DIRBE instrument measured $F_K = 121.7 \pm 11.4$\,Jy at $2.2\,\mu$m (Smith et al.~\cite{smith04}), corresponding to $m_K = 1.77 \pm 0.10$, and the 2MASS catalog (Skrutskie et al.~\cite{skrutskie06}) gives an apparent $K$ band magnitude of $m_K = 1.72 \pm 0.26$ for Aab+B, equivalent to $F_K = 127 \pm 34$\,Jy. Averaging these two measurements, we obtain $m_K = 1.76 \pm 0.10$ and consequently:
\begin{equation}
 m_K ({\rm Aab}) = 1.86 \pm 0.09
 \end{equation}
\begin{equation}
 m_K ({\rm B}) = 4.40 \pm 0.09
 \end{equation}

\subsection{Photometry during the primary eclipse}

The primary eclipse of Aab in the $K$ band observed with NACO can give us the difference in surface brightness of Ab and Aa. As this eclipse is not total, but annular, we cannot derive unambiguously the brightness ratio of the two stars as it depends on their angular diameters. In this paragraph, we consider that the eclipse is perfectly centered, thus neglecting a possible differential effect of the limb darkening (LD) of the two stars. As we observed very close to the eclipse minimum and in the $K$ band, where the LD is small (the LD/UD correction is less than 3\%), this appears as a reasonable assumption. We also neglect the possible distortion of the photospheres of Aa and Ab due to a possible fast rotation.

During the primary eclipse, the disk of Ab ``replaces" part of the disk of Aa with a lower effective temperature (hence lower surface brightness) ``patch". The eclipse depth in the $K$ band is $0.44\pm 0.01$\,mag (Sect.~\ref{nacophot}).
This can be compared with the depth of the primary eclipse in the visible: $\Delta m_V = 0.53 \pm 0.01$. The ``color" of the primary eclipse is therefore $\Delta m_V- \Delta m_K = 0.09 \pm 0.02$\,mag. This corresponds to the ``differential difference" in color between the two stars between the visible and IR wavelengths. This leads to a magnitude difference of:
\begin{equation}
m_K({\rm Ab}) - m_K({\rm Aa}) = 1.02 \pm 0.04
\end{equation}
corresponding to a flux ratio $f({\rm Ab})/f({\rm Aa})=39.2 \pm 1.4\,\%$ and:
\begin{equation}
 m_K ({\rm Aa}) = 2.22 \pm 0.09
 \end{equation}
\begin{equation}
 m_K ({\rm Ab}) = 3.24 \pm 0.09
 \end{equation}

\subsection{Emergence of Ab from the secondary eclipse \label{eclipse2}}

It is interesting to notice that our  2008-04-23 NACO observation has been obtained only about 30\,minutes after the end of the secondary (total) eclipse during which the $\delta$\,Vel~Aa component covered Ab. The apparent disks of the two stars are therefore ``touching" each other. Our measurement shows a slight variation of the photometric ratio $\rho = f(B)/f(A)$, as we measure $9.90 \pm 0.12\%$, compared to an average value over the eight other out-of-eclipse epochs of $9.63 \pm 0.05\%$. This means that the masking of part of the close environment of $\delta$\,Vel~Ab by Aa results in a slight loss of total flux of the Aab system.

Although of marginal statistical significance ($2.1\,\sigma$), this measurement is the maximum value of the out-of-eclipse ratios we obtained. It could be due to the presence of a circumstellar flux contributor of unknown nature in the immediate vicinity of $\delta$\,Vel~Ab, accounting for $0.27 \pm 0.13\%$ of the total flux of Aab. If Ab is a fast rotator, and its rotation axis is perpendicular to the orbital plane, one hypothesis could be that it is the signature of a circumstellar disk made of gas (Ae episode).

\subsection{Angular diameters, radii and effective temperatures \label{delvel_params}}

From our spatially resolved photometry of the three stars of $\delta$\,Vel (Aa, Ab and B), we can deduce their limb-darkened angular diameters $\theta_{\rm LD}$, photospheric radii $R$ and effective temperatures $T_{\rm eff}$ using the surface brightness-color (SBC) relations calibrated by Kervella et al.~(\cite{kervella04}) in $(V, V-K)$ and the {\it Hipparcos} parallax. The results are summarized in Table~\ref{properties_delvel}. This approach allows us to rely only on empirical SBC relations (calibrated by interferometric angular diameter measurements) to compute the parameters of the three stars. Another approach to determine these parameters would be to adjust directly synthetic spectra to the photometry. While the comparison with such model spectra is necessary to determine the IR excess (Sect.~\ref{discussion}), it has at this stage the disadvantage of making the derived parameters dependent on the choice of one particular numerical model library. For this reason, we prefer to use the empirical SBC relations. In any case, the good agreement between the model spectra and the measured photometry in the visible and near-IR domains shown in the bottom parts of Fig.~\ref{bi-BB} and \ref{tri-BB} indicates that both methods give consistent results.

The parameters of $\delta$\,Vel are in good agreement with the values deduced for the combined system A+B from spectroscopy by Allende Prieto \& Lambert~(\cite{allende99}): $T_{\rm eff} = 8700$\,K, $R = 3.9\,R_\odot$. The larger radius is naturally explained by the combination of the flux from Aa and Ab (and to a lesser extent of B), while the effective temperature is a weighted average of the values of the three stars. They also find an effective gravity of $\log g = 3.66$ and a mass of $M = 2.51\,M_\odot$. The reliability of these values is uncertain due to the effect of the orbital motion of Aab on the spectral line broadening. However, we do not confirm the large radii derived by Kellerer et al.~(\cite{kellerer07}) for Aa and Ab (6.0 and $3.3\,R_\odot$) from interferometric observations of the trio.

%___________________Table
\begin{table*}
\caption{Properties of $\delta$\,Vel~Aa, Ab and B derived from their $V$ and $K$ band magnitudes using the surface brightness-color relations from Kervella et al.~(\cite{kervella04}). The bolometric corrections for the $V$ band were taken from Bessell et al.~(\cite{bessell98}). The approximate masses of the three stars are discussed in Sect.~\ref{discussion}.}
\label{properties_delvel}
\begin{tabular}{lccccrcrcc}
\hline \hline
Star & $\theta_{\rm LD}$\,(mas) & Radius\,($R_\odot$) & $T_{\rm eff}$\,(K) & $M_V$ & BC$_V$ & M$_{\rm bol}$ & $L/L_\odot$ & Spect. & $M/M_\odot$ \\
\hline
\noalign{\smallskip}
Aa & $1.21 \pm 0.03$ & $3.17 \pm 0.08$ & $9\,000 \pm 400$ & $0.39 \pm 0.04$ & -$0.07$ & $0.32 \pm 0.04$ & $59.2 \pm 2.0$ & A2IV & $2.5 \pm 0.1$\\
Ab & $0.77 \pm 0.02$ & $2.02 \pm 0.05$ & $8\,600 \pm 350$ & $1.50 \pm 0.04$ & 0.00 & $1.50 \pm 0.04$ & $20.0 \pm 0.7$ & A4V & $2.0 \pm 0.1$\\
B & $0.53 \pm 0.02$ & $1.39 \pm 0.06$ & $6\,350 \pm 350$ & $3.60 \pm 0.04$ & 0.00 & $3.60 \pm 0.04$ & $3.1 \pm 0.2$ & F8V & $1.3 \pm 0.1$\\
 \hline
\end{tabular}
\end{table*}

%__________________________________Spectrum modeling
\section{Discussion \label{discussion}}

The IR excess detected around $\delta$\,Vel by IRAS was recently attributed by G\'asp\'ar et al.~(\cite{gaspar08}) to interstellar dust heated by $\delta$\,Vel. In this Section, we examine the possibility that one of the components of $\delta$\,Vel~A or B presents an IR excess of circumstellar origin. For this purpose, we compare the fluxes extracted from our spatially resolved VISIR observations to the expected photospheric flux from the Aab and B components.

\begin{figure}[]
\centering
\includegraphics[bb=0 0 360 288, width=8.7cm, angle=0]{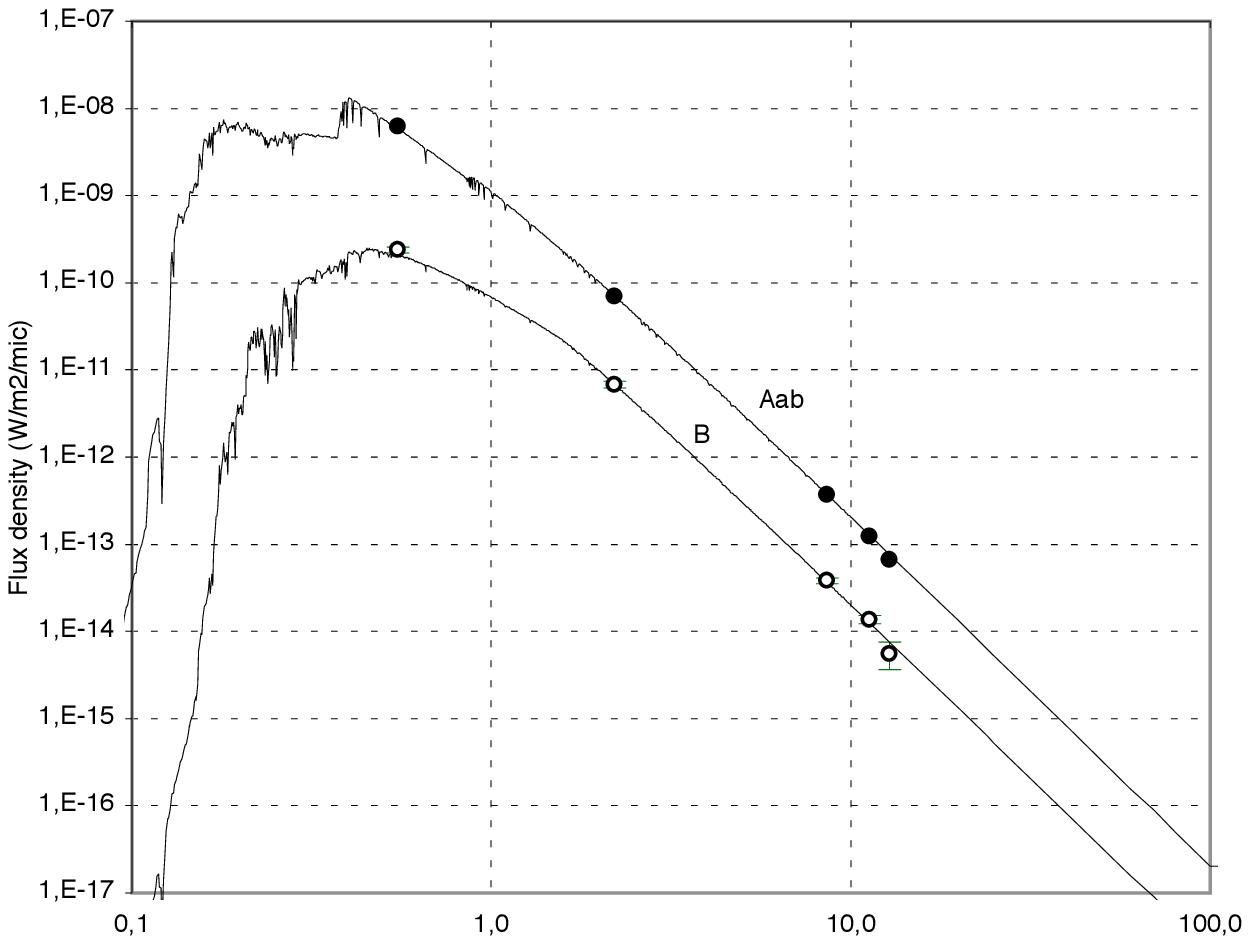}
\includegraphics[bb=0 0 360 144, width=8.7cm, angle=0]{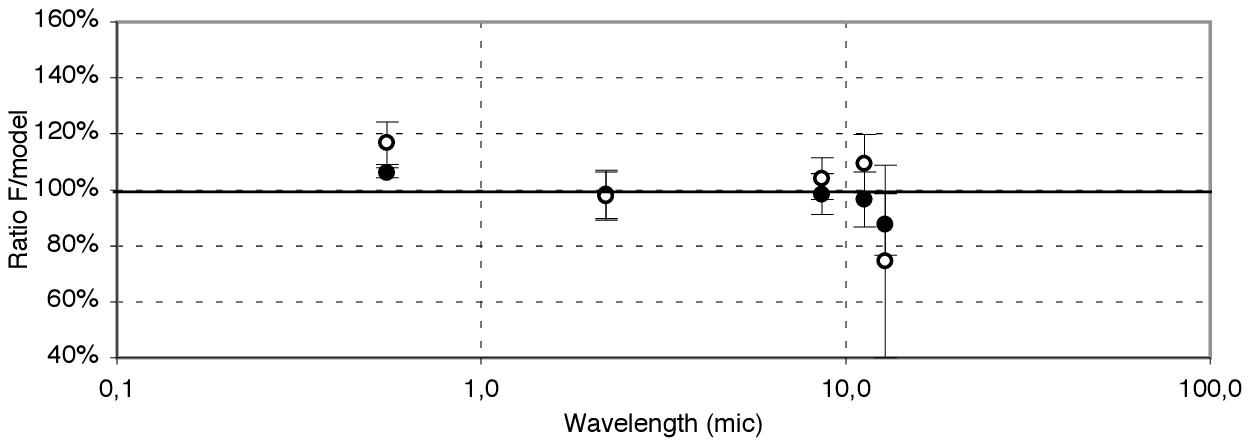}
\caption{Spatially resolved photometry of $\delta$\,Vel~Aab and B compared to Castelli \& Kurucz~(\cite{castelli03}) models of Aa+Ab and B, considering the stellar properties listed in Table~\ref{properties_delvel} (top) and the residuals compared the observations (bottom).  \label{bi-BB}}
\end{figure}

\begin{figure}[]
\centering
\includegraphics[bb=0 0 360 288, width=8.7cm, angle=0]{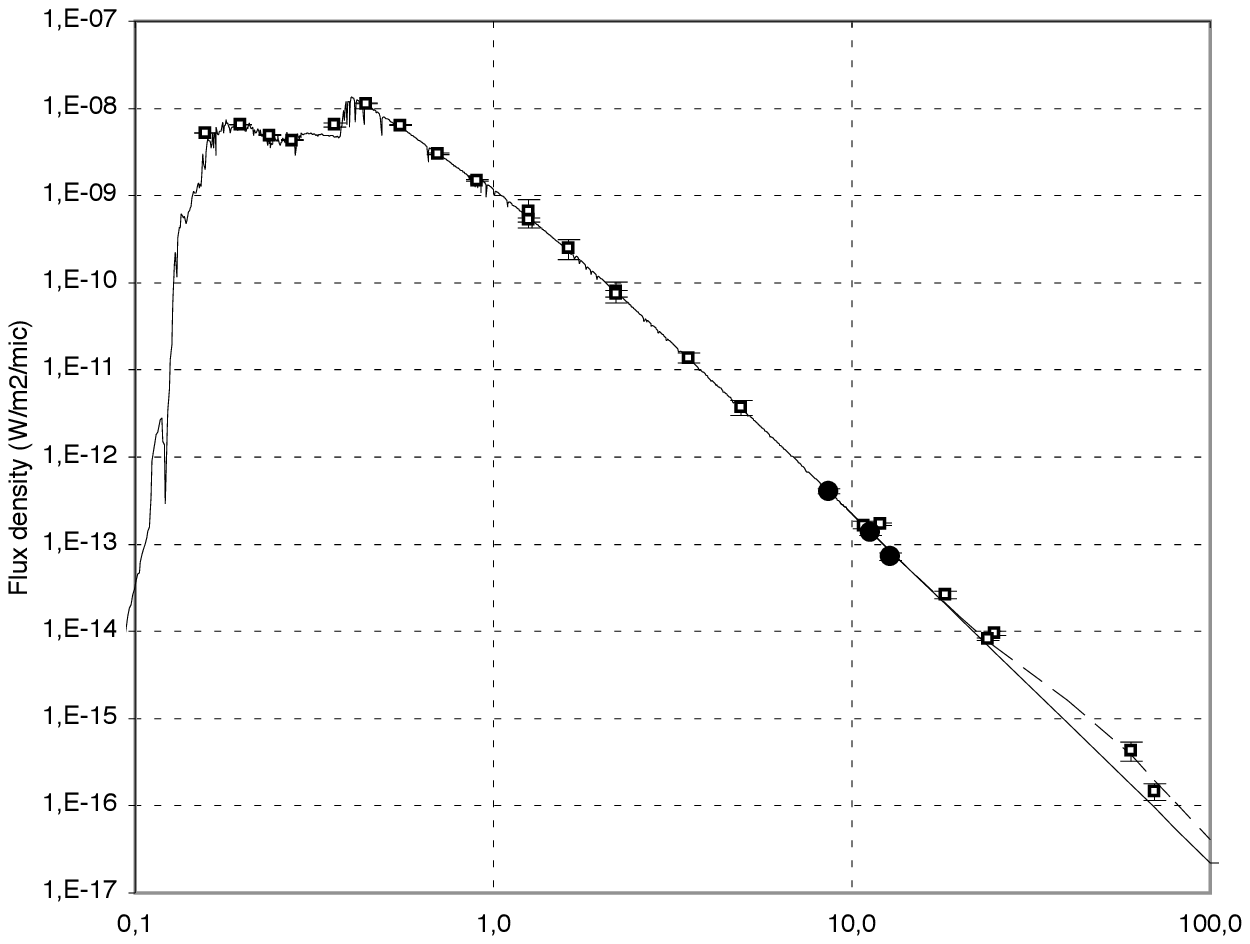}
\includegraphics[bb=0 0 360 144, width=8.7cm, angle=0]{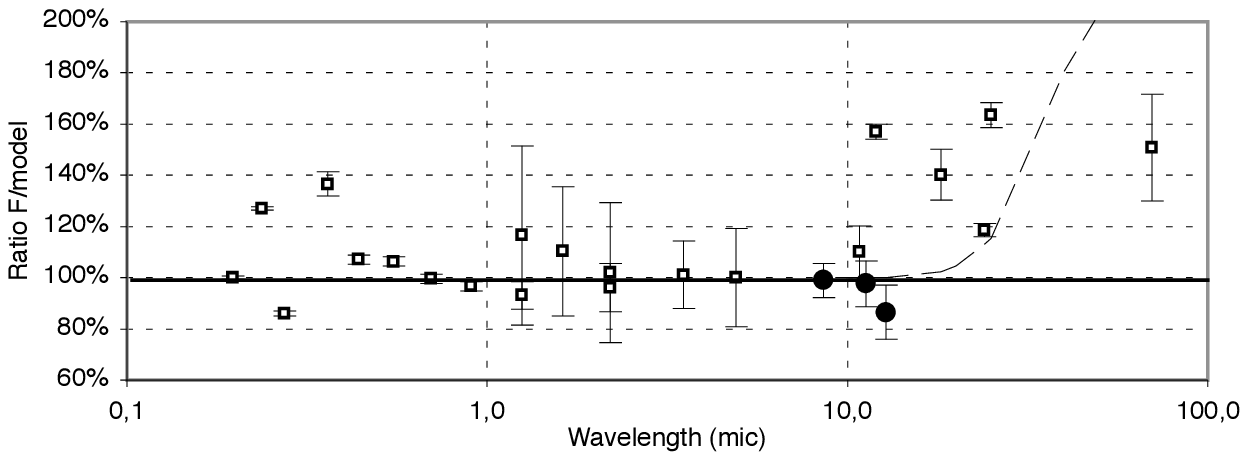}
\caption{Total model flux of $\delta$\,Vel Aab+B (solid curve), together with the available photometry from the literature (squares) and the VISIR photometry (filled circles). The dashed curve is the flux contribution of the IR excess model proposed by G\'asp\'ar et al.~(\cite{gaspar08}). The bottom plot shows the relative photometry with respect to the combined photospheric model of the three stars. \label{tri-BB}}
\end{figure}

%___________________Table
\begin{table}
\caption{Observed fluxes and excesses of $\delta$\,Vel compared to the Castelli \& Kurucz~(\cite{castelli03}) models. $N\sigma$ is the measured excess expressed in number of times the measurement uncertainty.}
\label{fluxes_delvel}
\begin{tabular}{lcccc}
\hline \hline
Band & $\lambda$ ($\mu$m) & $F$ ($10^{-13}$ W/m$^2$/$\mu$m)   & $F/F_{\rm mod}$ (\%) & $N\sigma$ \\
\hline
{\it Aab+B} \\
PAH1 & 8.59 & $ 4.07 \pm 0.27  $ & $ 99 \pm 7$ & -0.2 \\
PAH2 & 11.25 & $ 1.38 \pm 0.12 $ & $ 98 \pm 9$ & -0.3 \\
NeII & 12.81 & $ 7.26 \pm 0.08 $ & $ 86 \pm 11$ & -1.3 \\
\hline
{\it Aab} \\
PAH1 & 8.59 & $ 3.69 \pm 0.27  $ & $ 98 \pm 7$ & -0.2 \\
PAH2 & 11.25 & $ 1.24 \pm 0.12 $ & $ 97 \pm 10$ & -0.3 \\
NeII & 12.81 & $ 0.67 \pm 0.07 $ & $ 88 \pm 11$ & -1.1 \\
\hline
{\it B} \\
PAH1 & 8.59 & $ 0.382 \pm 0.028  $ & $ 104 \pm 7$ & 0.5 \\
PAH2 & 11.25 & $ 0.138 \pm 0.014 $ & $ 109 \pm 10$ & 0.9 \\
NeII & 12.81 & $ 0.056 \pm 0.019 $ & $ 74 \pm 34$ & -0.7 \\
 \hline
\end{tabular}
\end{table}

To retrieve the synthetic stellar spectra of the three stars from the library assembled by Castelli \& Kurucz~(\cite{castelli03}), hereafter CK03, we used the parameters listed in Table~\ref{properties_delvel}. We read the tables for $\log g$ values of 4.0 for $\delta$\,Vel~Aa and Ab, and 4.5 for B, and solar metallicities. The sensitivity of the models to small variations of these two parameters is minimal. We would like to emphasize that we did not fit these spectra to the available photometry, but we simply multiplied the CK03 spectra by the squared limb-darkened angular diameter of each star (Table~\ref{properties_delvel}).
From these spectra, we derived the expected thermal IR flux from the stellar photospheres of Aa+Ab and B in the PAH1, PAH2 and NeII filters corresponding to our VISIR observations.

We can also add the CK03 spectra of the three stars in order to compare the predicted flux of the trio to the existing non spatially resolved measurements of A+B from the literature. The resulting spectrum is shown in Fig.~\ref{tri-BB}. The photometric measurements are taken from Thompson et al.~(\cite{thompson78}; UV fluxes from TD1), Morel et al.~(\cite{morel78}; $UBVRI$), Skrutskie et al.~(\cite{skrutskie06}; 2MASS $JHK$), Smith et al.~(\cite{smith04}; COBE/DIRBE, four wavelengths from 1.25 to 4.5\,$\mu$m), VISIR (8.6, 11.25 and $12.81\,\mu$m, this work), Jayawardhana et al.~(\cite{jayawardhana01}; $N$ and $18.2\,\mu$m bands), G\'asp\'ar et al.~(\cite{gaspar08}; \emph{Spitzer}/MIPS 24 and 70\,$\mu$m), and IPAC~(\cite{ipac86}; IRAS 12, 25 and 60\,$\mu$m). The additional flux from the bow shock model proposed by G\'asp\'ar et al.~(\cite{gaspar08}) is shown on both panels of Fig.~\ref{tri-BB} as a dashed curve. Its contribution in the range of wavelengths sampled by our VISIR observations appears negligible ($\approx 0.1\%$ of the photospheric flux).

As shown in Fig.~\ref{bi-BB}, Fig.~\ref{tri-BB} and Table~\ref{fluxes_delvel}, the comparison with our VISIR spatially resolved photometry of $\delta$\,Vel~A and B does not show a significant circumstellar IR excess in the VISIR bands compared to the expected photospheric flux of each of the two components, at a few percent level. 
The excess that is visible longwards of $10\,\mu$m in Fig.~\ref{tri-BB} comes from the flux contribution of the ISM bow shock at large angular distances that has been observed by G\'asp\'ar et al.~(\cite{gaspar08}). We note that the IRAS and Jayawardhana et al.~(\cite{jayawardhana01}) photometry shows an excess compared to their model between 12 and 25\,$\mu$m. This may be a consequence of the fact that the flux contribution from the ISM is (at least partly) included in the large aperture measurements (IRAS in particular), but not in our narrow aperture photometry.

We notice a relatively large ultraviolet excess on $\delta$\,Vel in Fig.~\ref{tri-BB}. A fast rotating star usually presents a UV excess due to its overheated polar caps (von Zeipel effect, see e.g. Aufdenberg et al.~\cite{aufdenberg06}). A discussion on the rotational velocity of the components of $\delta$\,Vel~A can be found in Argyle, Alzner \& Horch~(\cite{argyle02}).  As a remark, Royer, Zorec \& Gomez~(\cite{royer07}) found a rotational velocity of $v \sin i = 150$\,km/s, but this was computed for the combined spectrum of the three stars of $\delta$\,Vel so a bias may be present.

\begin{figure}[]
\centering
\includegraphics[bb=0 0 360 288, width=8.7cm, angle=0]{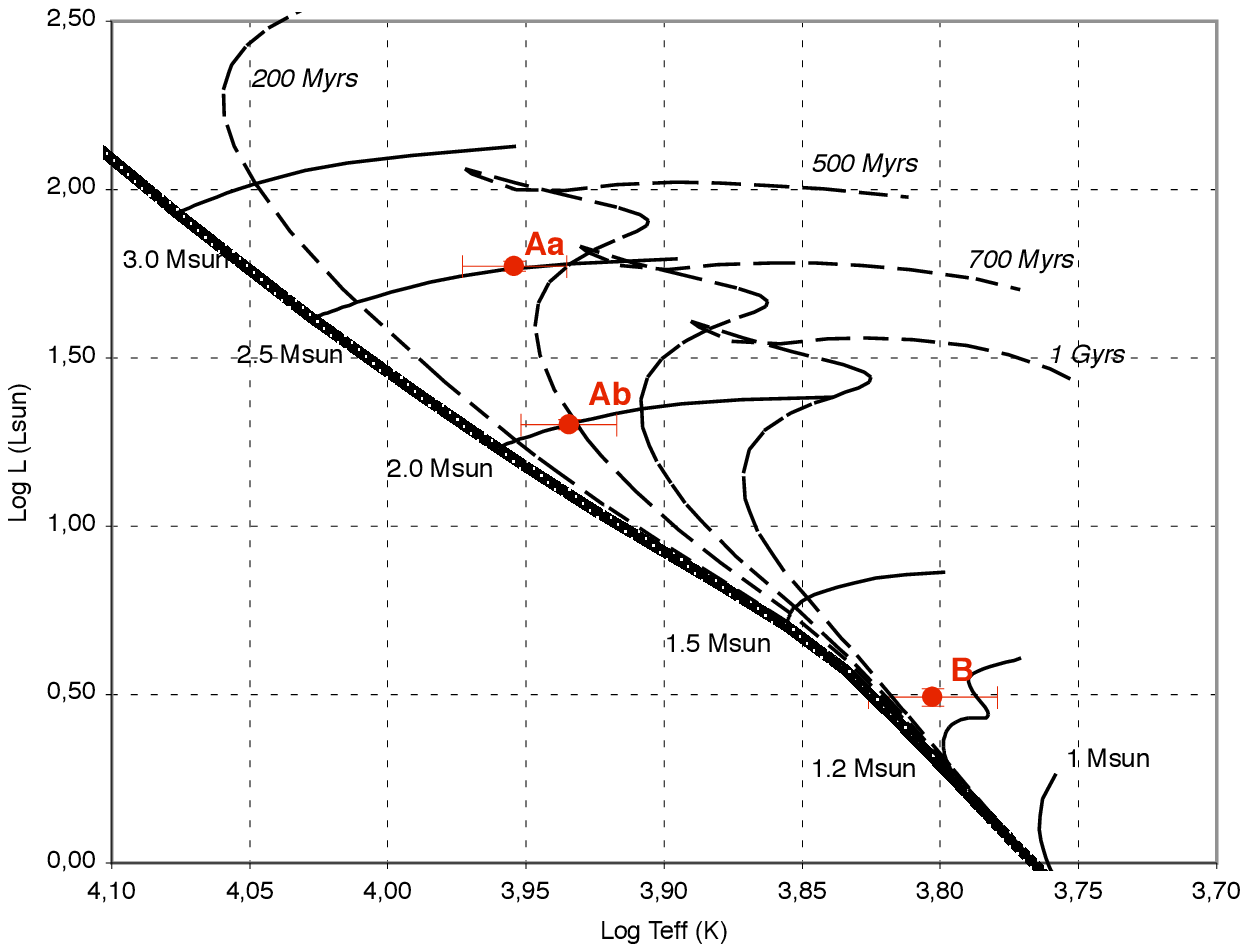}
\caption{Positions of $\delta$\,Vel Aa, Ab and B in the HR diagram, with the isochrones and isomass curves from Girardi et al.~(\cite{girardi00}) overplotted.  \label{HR-diagram}}
\end{figure}

The physical parameters derived for the three stars (Table~\ref{properties_delvel}) indicate that $\delta$\,Vel~Aa already evolved away from the main sequence. This is confirmed by the evolutionary tracks from Girardi et al.~(\cite{girardi00}), shown in Fig.~\ref{HR-diagram}, that give approximate masses of 2.5, 2.0 and 1.3\,$M_\odot$ respectively for Aa, Ab and B, and an age of approximately $400-500$\,Myrs for the system. This age compares well with the age of 330-390\,Myr proposed by Rieke et al.~(\cite{rieke05}), while the masses are in relatively good agreement with Argyle, Alzner \& Horch~(\cite{argyle02}), who determined a total dynamical mass of the $\delta$\,Vel\,AB system of $5.7^{+1.3}_{-1.1}\,M_{\odot}$.
We postpone a more detailed discussion of the ages and evolutionary status of the $\delta$\,Vel stars to a forthcoming paper.

%__________________________________Conclusion
\section{Conclusion}

Our VISIR photometry of $\delta$\,Vel A and B does not show the presence of an excess of circumstellar origin in the thermal IR domain ($8-13\,\mu$m) at a level of a few percents.  This result indicates that these stars do not host a warm debris disk with a typical temperature around 200-300\,K. This supports the conclusions of G\'asp\'ar et al.~(\cite{gaspar08}) who attribute the IR excess detected with \emph{Spitzer} to the surrounding interstellar medium material.
The absence of a warm circumstellar disk in the inner $\delta$\,Vel system may be a consequence of the gravitational interactions between the three stars. They could have caused the dispersion of the residual circumstellar material from which they stars formed. However, the possibility still exists that a cold debris disk is present, as in the case of Vega (Su et al.~\cite{su05}) and Fomalhaut (Stapelfeldt et al.~\cite{stapelfeldt04}), but at large distances from the stars. In order to test this scenario, observations of $\delta$\,Vel in the far IR or submillimetric domain similar for instance to those of Fomalhaut by Holland et al.~(\cite{holland03}) would be necessary.
%Considering the current angular separation of the two stars, this would require at least a 20\,m aperture at $\lambda = 70\,\mu$m.

From the flux ratios in the $V$ and $K$ bands, we could derive the physical properties of the three stars Aa, Ab and B. $\delta$\,Vel appears as a moderately evolved system, with the Aa component currently leaving the main sequence. As suggested by the $v \sin i$ value of Royer et al.~(\cite{royer07}), the ultraviolet excess of $\delta$\,Vel and our putative observation of near-IR circumstellar emission close to Ab may indicate that $\delta$\,Vel Ab (and/or Aa) could be a fast rotator, turning this eclipsing system into a promising object to map the polar brightening associated to the von Zeipel effect (von Zeipel~\cite{vonzeipel24}).

%__________________________________Acknowledgements
\begin{acknowledgements}
We are grateful to the ESO observing team at Paranal for the perfect execution of
our NACO observations of $\delta$\,Vel, in particular for the time-critical eclipse on May 18th, 2008.
We also warmly thank Sebastian Otero (Asociaci\'on Cielo Sur, Argentina),
Brian Fraser (Sunninghill Observatory, South Africa) and Jaime Garc\'ia
(Instituto Cop\'ernico, Argentina) for providing the visible photometry of $\delta$\,Vel.
This work received the support of PHASE, the high angular resolution partnership between
ONERA, Observatoire de Paris, CNRS and University Denis Diderot Paris 7.
This research took advantage of the SIMBAD and VIZIER databases at the CDS, Strasbourg 
(France), and NASA's Astrophysics Data System Bibliographic Services.
\end{acknowledgements}

%__________________________________Bibliography
{}

\end{document}